\documentclass[a4paper,titlepage]{article}
\usepackage{setspace}

\usepackage{arxiv}
\usepackage{cite}
\usepackage{amsmath,amssymb,amsfonts}
\usepackage{algorithmic}
\usepackage{graphicx}
\usepackage{tikz}
\usepackage[percent]{overpic}  
\usepackage{svg}
\usepackage{textcomp}
\usepackage{adjustbox}
\usepackage{xcolor}
\usepackage{hyperref}
\usepackage{listings}
\usepackage{tabularx}
\usepackage[T1]{fontenc}
\usepackage{minted}
\usepackage{subcaption}
\usepackage{pifont}  
\newcommand{\cmark}{{\color{green!60!black} \ding{52}}}
\newcommand{\xmark}{\color{red} \ding{56}} 
\def\BibTeX{{\rm B\kern-.05em{\sc i\kern-.025em b}\kern-.08em
T\kern-.1667em\lower.7ex\hbox{E}\kern-.125emX}}
\lstdefinelanguage{qasm}{
    morekeywords={OPENQASM, qreg, creg, measure, cx, u3, u2, u1, barrier},
    sensitive=true,
    morecomment=[l]{//},
    morestring=[b]"
}

\lstdefinestyle{pythoncode}{
    basicstyle=\ttfamily\small,
    keywordstyle=\color{blue}\bfseries,
    commentstyle=\color{gray}\itshape,
    stringstyle=\color{red},
    numbers=left,
    numberstyle=\tiny,
    stepnumber=1,
    numbersep=5pt,
    breaklines=true,
    showstringspaces=false
}

\begin{document}

    \title{Qompiler: A Traceable Quantum Circuit Synthesizer for Arbitrary Hamiltonians}

    \author{Shoupu Wan \\
    Email: wanshoupu@gmail.com \thanks{Github link: \url{https://github.com/wanshoupu/quantum-simulations}}
    }

    \maketitle

    \begin{abstract}
        We present a quantum compiler framework that bridges the gap between physics modeling and quantum software development.
        At the core of this framework is a versatile quantum circuit synthesizer capable of decomposing arbitrary Hamiltonians into quantum circuits,
        represented using a platform-independent B-Tree-based intermediate representation.
        The B-Tree structure encodes information for gate lineage, enabling detailed tracing information of quantum circuit gates and facilitating circuit verification.
        The intermediate representation serves as a universal, hardware-agnostic carrier of compiled code,
        allowing it to be readily rendered on most quantum hardware backends and transpiled into other quantum circuit languages.
        We demonstrate rendering the intermediate representation into executable quantum circuits in Qiskit and Cirq.
        We can also transpile the intermediate representation into OpenQASM for broader compatibility.
    \end{abstract}


    \section{Introduction}\label{sec:introduction}

    Quantum simulation offers significant advantages in fields such as chemistry, pharmaceuticals,  biomedicine, and materials science.
    Given a Hamiltonian $\hat{H}$, the time-dependent evolution operator can be expressed as
    \begin{equation}
        \label{eq:hamiltonian}
        \hat{U}\left( t \right) = e^{-i\hat{H}t/\hbar},
    \end{equation}
    where $t$ is time and $\hbar$ is Planck's constant.
    In Schrödinger picture, the state ket of the system as a function of time is
    \begin{equation}
        \label{eq:sysstate}
        \left\| \psi\left( t \right) \right\rangle = \hat{U}\left( t \right)\left\| \psi_{0} \right\rangle
    \end{equation}
    where $\left\| \psi_{0} \right\rangle$ is the initial ket of a system state.
    Equivalently in Heisenberg picture, the observables are time-dependent, namely,
    \begin{equation}
        \label{eq:heisenberg}
        \hat{A} \left( t \right) = \hat{U}^{\dagger} \left( t \right) \hat{A}_{0} \hat{U} \left( t \right)
    \end{equation}
    where $\hat{A}_{0}$ is a constant operator for the observable at time $t=0$.
    There are other presentation pictures than these two.
    Regardless of picture, the quantum dynamics of the system are fully encoded in $\hat{U}\left( t \right)$.
    Therefore, as long as we can reproduce the operator $\hat{U}\left( t \right)$ in a quantum circuit, we can simulate the dynamics of the system.
    In this article, we assume the operator $\hat{U}\left( t \right)$ has been evaluated as time series of unitary matrices,
    $\left\{ U_{i} |i = 0, 1, \dots, n \right\}$ on a suitable basis has chosen in the system Hilbert space by Troterization
    or other equivalent methods~\cite{Suzuki1991,AbramsLloyd1999QPE,Lloyd1996,Chuang1997QPT}.

    Despite its potential in quantum simulation, writing full-stack quantum computing software requires domain knowledge
    in physics, mathematics, computer science, etc.
    The steep learning curve poses a barrier and limits access for many potential users and professionals
    who could benefit from quantum computing.
    Existing quantum transpilers are often platform-specific, implemented in an ad hoc manner,
    and offer limited control.
    There is clearly a need of a software system to fill the gap.

    We hereby present \texttt{Qompiler}, a cross-platform quantum compiler
    featuring automation, scalability, traceability, verifiability, and configurability that we attempt to address the abovementioned challenges.
    \texttt{Qompiler} decomposes the evolution operator, $\hat{U}$, into quantum circuits
    using multi-stage decomposition processes.
    The end result (quantum circuit) is saved into an
    intermediate representation (IR), a B-Tree-based data structure.
    This IR is the key for platform independence because out of it we can generate quantum circuits across multiple quantum platforms with varying hardware backend
    through polymorphically implementing the base class \texttt{CircuitBuilder}.
    We have realized support on platforms including Qiskit, Cirq, and OpenQASM~\cite{cirq2024,qiskit2023,openqasm2,openqasm3}.
    Notably, \texttt{Qompiler}'s powerful configuration capabilities also enable fine-grained gate selection
    that aligns with the native hardware.
    For example, on superconducting hardware, standard Pauli gates may be preferable than more complicated gates,
    whereas on trapped-ion devices, parametric rotation gates can be more efficiently implemented.
    This hardware-aware optimization is designed into the foundation of the \texttt{Qompiler}.

    Table~\ref{tab:feature-comparison} compares \texttt{Qompiler} with existing quantum compilers,
    transpilers, and synthesizers according to the following criteria: cross-platform support,
    gate traceability, and compilation control.
    \texttt{Qompiler} stands out due to its combination of traceability, platform agnosticism, and configurable granularity control,
    as it will be shown in the following sections.
    \begin{table}[tbhp]
        \centering
        \caption{Comparing \texttt{Qompiler} and existing quantum compilation tools~\cite{sivarajah2020tket,smith2020quilc,younis2020qfast,amy2019staq}.
        Green check marks indicate support and red cross marks indicate lack of support.
            (A gray check mark is associated with Qiskit for its partial control over gate types.)
        }\label{tab:feature-comparison}
        \begin{tabularx}{.8\textwidth}{X|c|c|c|c|c|c|c}
            \hline
            \textbf{Feature}          & \textbf{Qiskit} & \textbf{Tket} & \textbf{Cirq} & \textbf{Quilc} & \textbf{QFast} & \textbf{Staq} & \textbf{Qompiler} \\
            \hline
            Cross-platform support         & \cmark          & \cmark        & \cmark        & \xmark         & \xmark         & \cmark        & \cmark \\
            Gate traceability & \xmark          & \xmark        & \xmark        & \xmark         & \xmark         & \xmark        & \cmark \\
            Gate granularity control    & \textcolor{gray}{\ding{52}} & \xmark & \xmark & \xmark & \cmark & \xmark & \cmark \\
            \hline
        \end{tabularx}
    \end{table}

    This compiler may be integrated with software stacks like Munich Quantum Software Stack~\cite{}.

    \section{High-Level Design}\label{sec:high-level-design}

    \begin{figure}[tbhp]
        \centering
        \includegraphics[width=.8\linewidth]{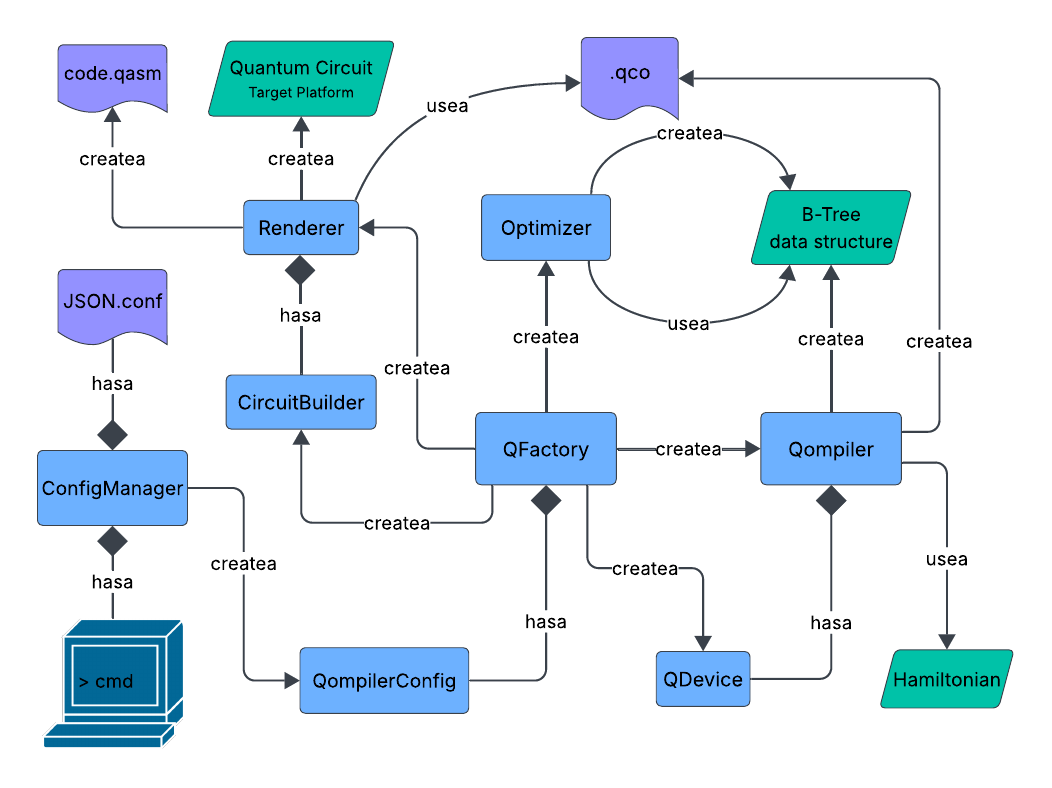}
        \caption{The high-level design of the `\texttt{Qompiler}' system.
        Rectangles (blue) represent major class constructs.
        Rectangles with a jagged bottom (purple) represent file input/output (file I/O).
        Parallelograms (pink) represent in-memory data structure.
        The computer icon represents user input.
        The lines specify the relationships between the entities.
        The relationships are one of the following---\texttt{hasa}, \texttt{usea}, \texttt{createa}.
        }
        \label{fig:impl}
    \end{figure}

    Let's go over the high-level design of \texttt{Qompiler}.
    The major functionalities of the system are organized into the following modules:
    ``Configuration'', ``Decomposition'', ``Optimization'', and ``Rendering'' (See Figure~\ref{fig:impl}).
    We will briefly describe each of these modules here, while leaving detailed discussions of some of these to later sections.

    Rich controls of \texttt{Qompiler} are made possible through the sophisticated configuration module,
    offering control over almost all aspects of the compilation process,
    including but not limited to the granularity control, target platform, error tolerances,
    quantum space allocation, optimization level, etc.
    This makes it possible for users to precisely tailor the resulting quantum circuit
    to best suit the targeted hardware.
    For instance, superconducting quantum hardware often implements Pauli gates ($X$, $Y$, $Z$) virtually,
    offering zero-error and zero-latency execution;
    So for superconducting quantum hardware, we can configure the compiler to prioritize Pauli gates.
    On the other hand, trapped-ion platforms natively support parametric rotation operators
    like $R_x(\phi)$, $R_y(\phi)$, and $R_z(\phi)$, and the \texttt{Qompiler} can be configured accordingly
    to produce rotation-based circuits.
    This hardware-aware configurability enables efficient, optimized circuit generation across diverse quantum backends
    while maintaining traceability and cross-platform compatibility.

    The configuration module is centered around the class \texttt{ConfigManager}.
    \texttt{ConfigManager} can read inputs from several sources: default settings, custom overrides (both JSON files),
    and command line settings.
    Once all configuration sources are loaded, \texttt{ConfigManager} combines and creates an overall configuration object
    which is used to create a singleton instance of \texttt{QFactory}.
    From the \texttt{QFactory}, other compilation constructs are created as
    needed---\texttt{Qompiler}, \texttt{QRenderer}, \texttt{QDevice}, \texttt{QConfig}, \texttt{Optimizers}, etc.
    All of these consciously follow the factory design pattern~\cite{Gamma1994DesignPattern}.

    Now that all the needed module objects are in place, \texttt{Qompiler} can start to process inputs.
    During the compilation process, the input operator undergoes a number of decomposition processes (see Table~\ref{tab:decomp}) recursively
    until the quantum gates reach the intended granularity, and all these are orchestrated by the \texttt{Qompiler} class
    as will be discussed in Section~\ref{sec:decomp}.
    The compilation output is a B-Tree data structure---IR\@.
    Beyond storing gate information and qubit interaction specifications,
    IR also tracks the parent--child relationships among the tree nodes, which is the root of
    the verifiability of quantum circuit to be discussed in Section~\ref{subsec:traceability-and-verifiability}.

    Before, we go into the details of the design of optimizer classes, some additional words would help understand the rationale.
    In effect, decomposition resembles disintegration, breaking the circuit into finer components,
    while the optimization process acts like integration, undoing the undesirable effect of decomposition.
    These complementary processes work like opposing push-and-pull forces to provide us precise control over gate granularity
    and help achieve an optimal balance in the compilated code that best suits the targeted quantum hardware.

    Following the decomposition processes are the recombination passes.
    The abovementioned decomposition process resembles disintegration, ensuring that all resulting gates meet \emph{or exceed} the target granularity.
    During the process, redundant gates and gates with exceedingly fine granularity are inevitably generated and become sources of inefficiency.
    We need a process to `undo' the overly fragmented gates, and this is done through the recombination passes.
    Think of the decomposition processes as pushes while the recombination processes as pulls.
    With both pushes and pulls, we achieve precise control over gate granularity.
    During the recombination passes, gates within a sliding window of certain size are consolidated or annihilated wherever possible.
    Figures~\ref{fig:unoptimized-circuit} and~\ref{fig:optimized-circuit} illustrate the effect of the recombination passes side-by-side.
    The design will be presented in Subsection~\ref{subsec:optimization} with more details.

    Once all the decomposition and recombination steps are finished, the IR is serialized in
    binary code and written to a \texttt{.qco} file for the rendering stages.
    At the rendering stage, the \texttt{.qco} file is loaded from the disk back into the B-tree data structure.
    Some additional optimization passes pertaining to the specific quantum computing backend may be performed.
    Note that the compute-intensive, weight-lifting compilation is done only once.
    Once we have the IR, circuit may be rendered as many times as needed for supported platforms.
    The design of circuit rendering is highly adaptable and open to new platforms.
    We will visit this in Section~\ref{subsec:rendering}.

    \section{Elementary Class Designs}\label{sec:design-of-elements}

    \begin{figure}[tbhp]
        \centering
        \includegraphics[width=.6\linewidth]{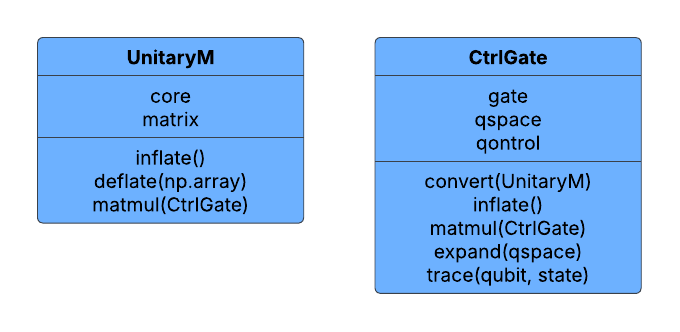}
        \caption{The class diagrams for \texttt{UnitaryM} and \texttt{CtrlGate}.}
        \label{fig:um-and-cgate}
    \end{figure}

    In this section, let's first review some key elementary classes of the \texttt{Qompiler} system.
    Among them are the classes for unitary matrices, quantum gates, IR, recombiners, renderer and circuit builders.

    \subsection{Unitary Matrices}\label{subsec:unitary-matrices}

    A multidimensional unitary matrix captures a snapshot of quantum dynamics of a specific system.
    It therefore has no reference to qubits in the simulation space yet.
    We created the class \texttt{UnitaryM} to represent a unitary matrix with two primary considerations.
    First, it encapsulates many good operations.
    It has a \texttt{matmul} method for matrix multiplication operations between \texttt{UnitaryM}'s,
    It also has a \texttt{deflate} method to convert a NumPy array ($2$D) into \texttt{UnitaryM},
    The \texttt{inflate} and \texttt{numpy.array} are created to convert \texttt{UnitaryM} objects back to NumPy arrays.
    Second, it provides an efficient framework for handling sparse matrices.
    For some use cases, the evolution operator $\hat{U}\left( t \right)$ is sparse, e.g.,
    the majority of the off-diagonal elements are zeros.
    Accordingly, we may store only the rows and columns with non-zero off-diagonal elements.
    We refer to the submatrix corresponding to these rows and columns the ``core-matrix''
    and the corresponding indices the ``core-indices''.
    Figure~\ref{fig:um-and-cgate} shows the class diagram for \texttt{UnitaryM}.

    \subsection{Quantum Space}\label{subsec:quantum-space}

    \begin{table}[htbp]
        \centering
        \begin{tabularx}{\linewidth}{l|X}
            \hline
            \textbf{Qubit type (enum \texttt{QType})} & \textbf{Description} \\
            \hline
            \texttt{TARGET} & A target qubit is the qubit on which a quantum gate performs an operation, namely, its quantum state may be altered as a result. \\
            \texttt{CONTROL0} / \texttt{CONTROL1} & A control qubit is one whose state determines whether an operation is applied to another qubit (\texttt{TARGET})
            while its own state is intact.
            A \texttt{CONTROL0} (\texttt{CONTROL1}) qubit is a control qubit with an activation value $0$ ($1$). \\
            \texttt{IDLER} & A qubit that does not participate in the gate operation but may appears in the operator matrix as identity elements. \\
            \hline
        \end{tabularx}
        \caption{Enum type \texttt{QType} for the roles of qubits in a particular quantum gate.}
        \label{tab:gate-role}
    \end{table}

    In a quantum circuit, there are two types of qubits according to the roles they played: data qubits and ancilla qubits.
    Data qubits are the working qubits that carry the computational states to be measured.
    Ancilla qubits are auxiliary qubits introduced to assist a quantum computation (e.g., for temporary storage, gate construction, or error correction)
    but not part of the final output, usually initialized in the fiducial state $\left\| 0 \right\rangle$.
    We also have a light-weight \texttt{Qubit} class for convenience where it makes distinction between ancilla and data qubits.
    We refer to the group of qubits involved in the quantum computation collectively as ``quantum space,'' or \texttt{qspace} for short.

    A quantum gate may involve more than one qubits.
    For example, in a \texttt{CNOT} gate, there is a control qubit and a target qubit.
    So another way to classify qubits is by the roles they play in a particular quantum gate.
    We create an \texttt{enum QType} for the roles played in individual quantum gates as shown in Table~\ref{tab:gate-role}.
    Note that we also introduce \texttt{IDLER} to denote that a qubit does not play a role in a gate operation and are simply present in the operator formulation as degrees of
    freedom.

    \subsection{Gate Granularity}\label{subsec:granularity}

    \begin{table}[htbp]
        \centering
        \caption{Terminology for granularity types (\texttt{enum GateGrain}) of quantum gates, in increasing order of fineness.
        When a gate meets the requirements of multiple categories, the finest one is chosen.
        For example, if a single-target gate has one control qubit a core matrix \texttt{X},
            we classify it as \texttt{Clifford+T}.
        }
        \begin{tabularx}{.9\textwidth}{l|X|l}
            \hline
            \textbf{Term} & \textbf{Description}                                                           & \textbf{Enum GateGrain}\\
            \hline
            Unitary       & Any $n$-qubit unitary operator                                                 & \texttt{UNITARY}       \\
            2level        & Unitary operator with no more than two non-identity rows/columns               & \texttt{TWO\_LEVEL}    \\
            Multi-Target  & Gates with $2+$ target qubit and any number of control qubits                  & \texttt{MULTI\_TARGET} \\
            Singlet       & Gates with a single target and any number of control qubits                    & \texttt{SINGLET}       \\
            Ctrl-pruned   & Controlled gates with a single target and at most one control qubit            & \texttt{CTRL\_PRUNED}  \\
            Principal     & Single-qubit gates of the form $R_n(\theta)$with $n=x,y,z$                     & \texttt{PRINCIPAL}     \\
            Universal     & Single-qubit gates in the set $\{X, Y, Z, H, S, T, \mathit{SD}, \mathit{TD}\}$ & \texttt{UNIV\_GATE}    \\
            Clifford+T    & Gates in the set $\{\mathit{CNOT}, H, S, T\}$                                  & \texttt{CLIFFORD\_T}   \\
            \hline
        \end{tabularx}
        \label{tab:terminology}
    \end{table}

    We introduce the notion of `granularity' and enum type \texttt{GateGrain} to classify quantum gates based roughly on the following factors:
    \begin{itemize}
        \item The number of qubits involved
        \item The number of \texttt{TARGET} qubits and the number of \texttt{CONTROL0}/\texttt{CONTROL0} qubits
        \item The operator features, e.g., Pauli operators, parametric rotation operator, \texttt{Clifford+T} operator, etc.
    \end{itemize}
    We listed the gate granularity types along with the enum values in Table~\ref{tab:terminology} for ease of reference.

    \subsection{Quantum Gates}\label{subsec:quantum-gates}

    Now we are ready to talk about design for quantum gates.
    A quantum gate is the basic building block of a quantum circuit that performs a unitary operation involving one or more qubits.
    So a quantum gate is necessarily tied up with a subspace in the simulation domain.
    We created the class `\texttt{CtrlGate}' as a general representation for quantum gates.
    It captures all the aspects of a quantum gates---the \texttt{qspace} it operates on, the roles played by each qubit in the \texttt{qspace},
    the core unitary operator to be applied to the target qubits.
    Therefore, \texttt{CtrlGate} can represent all types of quantum gate---single qubit gate, multi-qubit gates,
    single-target gates, multi-target gates, gates with or without control qubits, etc.

    For the core operator, \texttt{CtrlGate} can be constructed from Pauli operators $X$, $Y$, $Z$, Hadamard operator $H$, phase operators $S$, $T$
    and their Hermitian conjugates, parametric rotational operators $R_x(\phi)$, $R_y(\phi)$, and $R_z(\phi)$, or generic multi-target unitary matrices.
    As a good practice for object-oriented-programming, the \texttt{CtrlGate} class encapsulates the workhorse of most operations concerning a quantum gate:
    for matrix multiplications `\texttt{matmul}', for Hilbert space expansion \texttt{expand}, for sorting qubit \texttt{sorted},
    for trace operation \texttt{trace}, for direct conversion of \texttt{UnitaryM} into \texttt{CtrlGate} \texttt{convert}, etc.

    As keen readers have noticed, \texttt{IDLER} qubits are degrees of freedom that can be added into
    or taken away from the formulation of \texttt{CtrlGate} with not effect on the gate operation whatsoever.
    Expanding into additional Hilbert space is needed for verification when multiplying multi-qubit operators that span different qubit spaces.
    Adding extra qubits to \texttt{CtrlGate} is implemented by the method \texttt{expand}.
    By performing successive Kronecker product operations or Tracy-Singh matrix multiplication~\cite{TracySingh1972}, one can inject idler qubits in between the original qubit spaces.

    Another operation, useful especially for circuit verification, is the \texttt{trace} method.
    Because of the use of ancilla qubits, the \texttt{qspace} of the resulting circuit is expanded upon that of the input data qubits.
    We therefore need to eliminate these qubits from the multi-qubit gates in order to do direct comparisons with the input gate.
    The \texttt{trace} method comes handy for this.
    The class diagram for \texttt{CtrlGate} is also included in Figure~\ref{fig:um-and-cgate}.

    \subsection{ByteCode and Verifiability}\label{subsec:traceability-and-verifiability}

    As quantum circuits grow in size, verifiability becomes quite troublesome.
    Even a relatively simple molecule like LiH in the STO-3G basis would translate to a quantum circuit
    comprising approximately $1000$ gates~\cite{Hong_2022}.
    Simulation of moderately large molecules can consist of millions of gates~\cite{poulin2014,google_science_2020}.
    Keeping track of all these gates and ensuring correctness is challenging.

    \begin{figure}[tbhp]
        \centering
        \begin{overpic}[width=0.6\textwidth]{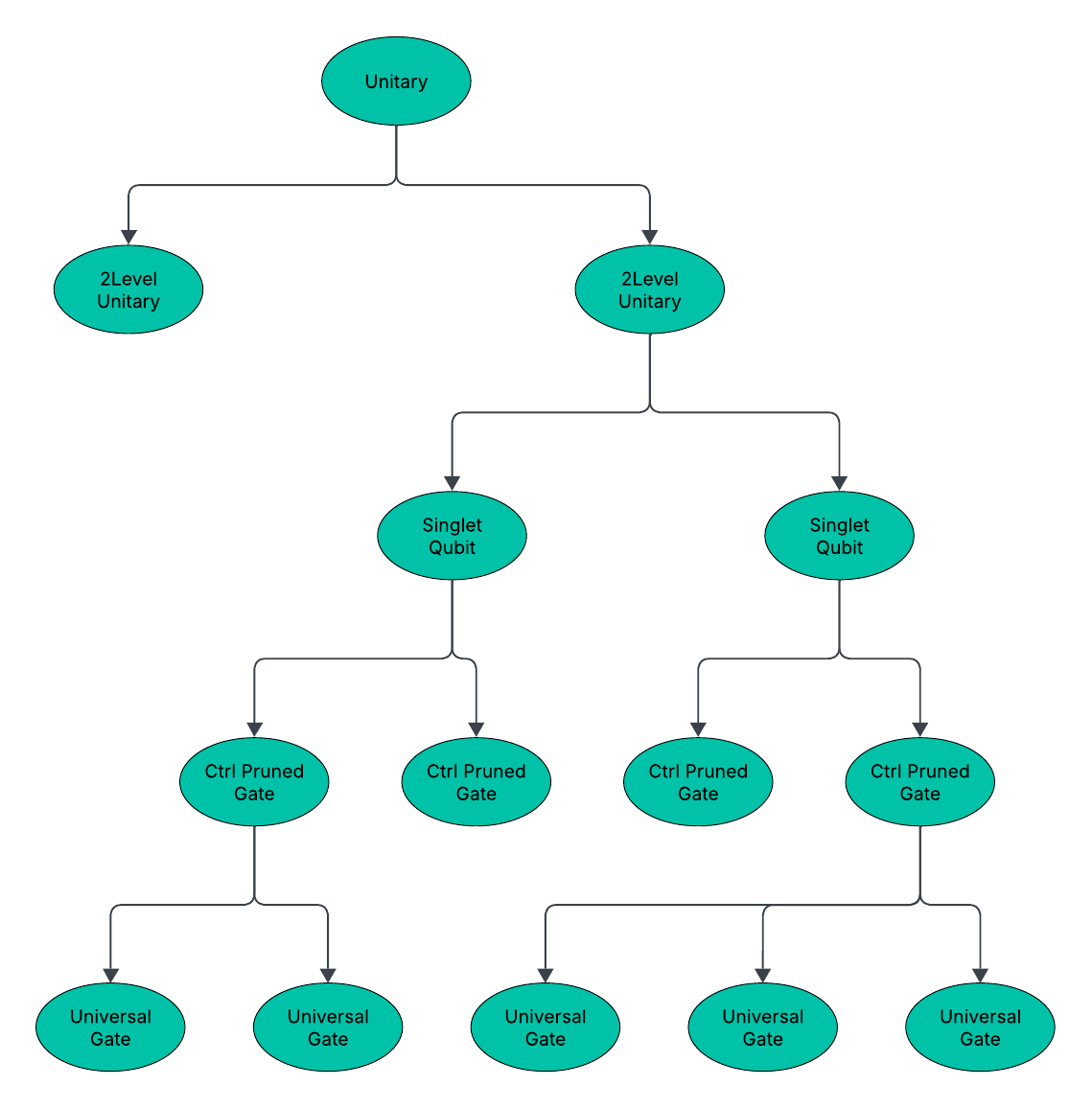}
            \put(0,70){\colorbox{white}{\makebox(20,10){}}}
        \end{overpic}
        \caption{The layered structure of, \texttt{ByteCode}, B-Tree data structure.
        The root of the tree is the input unitary matrix,
            followed by intermediate layers: two-level unitary, single-qubit, control-pruned.
            At the leave layer, are the components of the quantum circuit.
        }
        \label{fig:btree}
    \end{figure}

    A highlighted feature of our system is the built-in traceability and verifiability.
    Let's examine how it is achieved.
    To store and organize the compiled code as IR, we designed a B-Tree data structure called \texttt{ByteCode}
    where each node represents a compilation unit, a quantum operator, along with some associated metadata.
    Traceability is encoded in the parent--child relationships:
    each gate in the final program can be traced back to the specific operation that produced it.
    The built-in optimizers applied in later stages of compilation also record tracing information for their operations,
    preserving the original lineage.

    In the B-Tree data structure, parent nodes generally correspond to coarser-grained operators,
    whereas child nodes are finer-grained operators whose matrix product recovers the parent node, e.g.,
    \begin{equation}
        \label{eq:parent-child-relation}
        \hat{P} = \prod_{i=1}^{n} \hat{C}_{i}.
    \end{equation}
    Figure~\ref{fig:btree} illustrates the layered nature of the \texttt{ByteCode} data structure.

    Convenient methods such as \texttt{is\_leaf} and \texttt{herm} (Hermitian conjugate) are provided
    to simplify programming with tree nodes.
    Two iterators, \texttt{PreorderIterator} and \texttt{ReversePreorderIterator}, are also implemented
    to traverse the tree.
    In addition, we support serialization/deserialization (\texttt{serde}) of this data structure,
    to faithfully store and load the B-Tree data structure to and from \textit{.qco} files.

    With the tracking and tracing information of the \texttt{ByteCode} data structure by design, verification of the quantum circuit
    and program debugging is made easier in \texttt{Qompiler}.

    \subsection{Sliding Window Combiners}\label{subsec:optimization}

    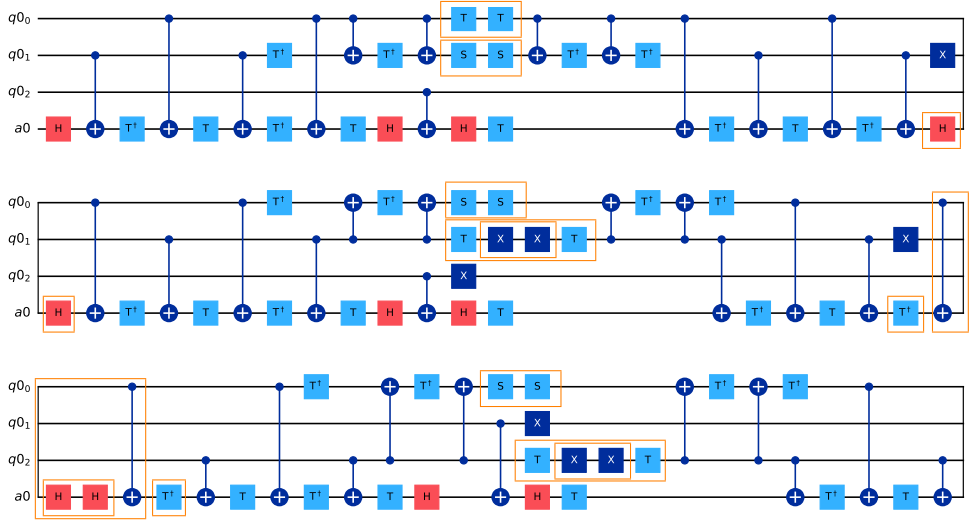
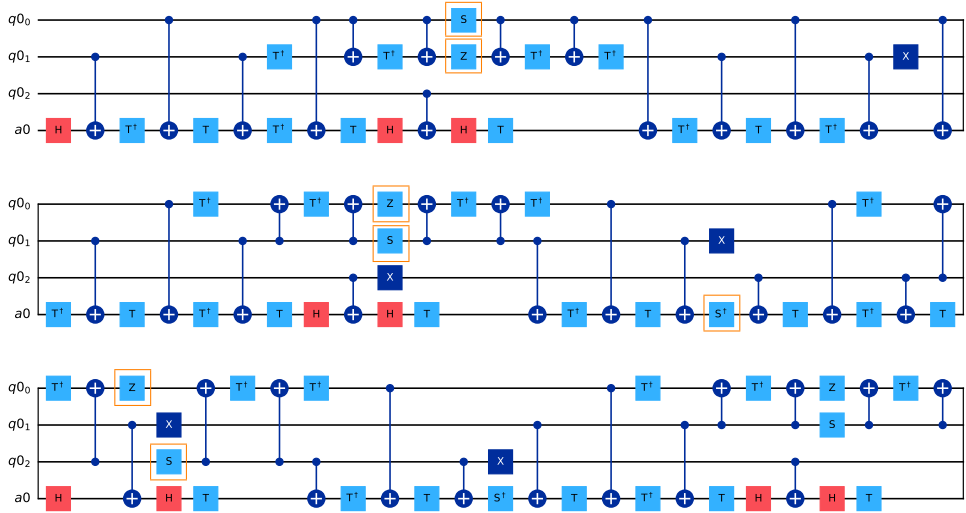
\begin{figure}[h]
        \centering
        \begin{subfigure}{0.8\textwidth}
            \centering
            \begin{overpic}[width=\linewidth,clip,trim=0 3000 0 30]{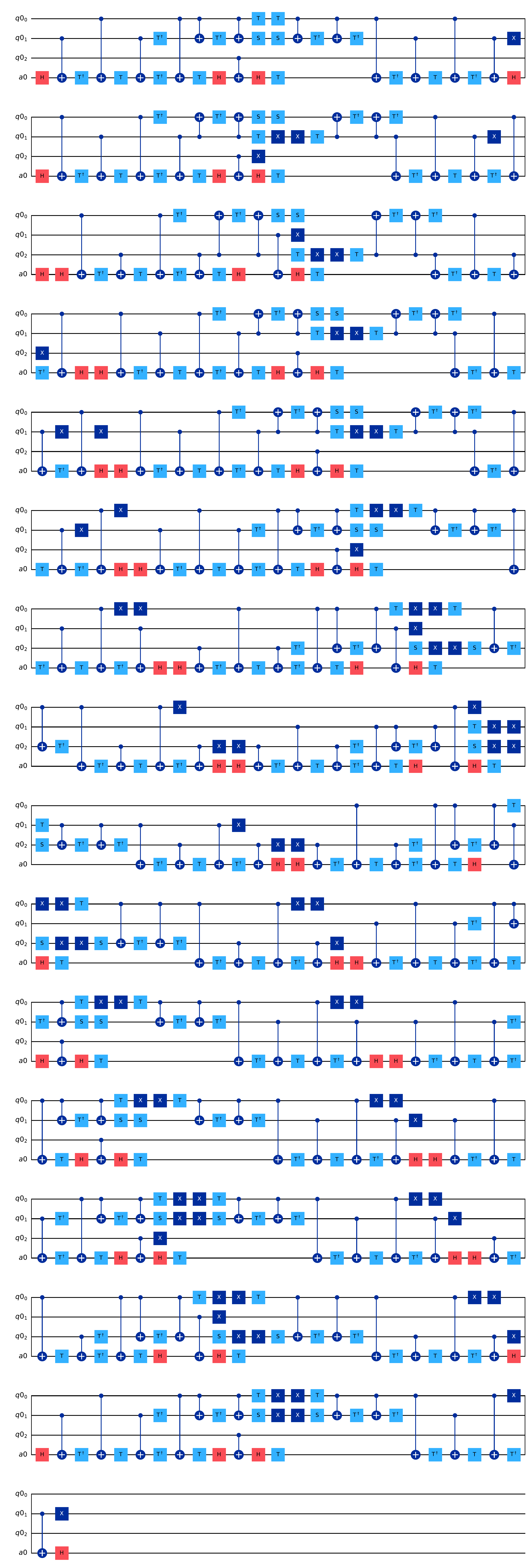}
                \put(46.5,50.2){\color{orange}\framebox(8,3.5){}}  

                \put(46.5,46.3){\color{orange}\framebox(8,3.5){}}  

                \put(47,32){\color{orange}\framebox(8,3.5){}}  

                \put(95,39){\color{orange}\framebox(3.7,3.5){}}  
                \put(6.5,20.5){\color{orange}\framebox(3,3.5){}}  

                \put(50.5,28){\color{orange}\framebox(7.5,3.5){}}  
                \put(47,27.7){\color{orange}\framebox(15,4){}}  

                \put(50.5,13){\color{orange}\framebox(8,3.5){}}  

                \put(58,5.7){\color{orange}\framebox(7.5,3.5){}}  
                \put(54,5.5){\color{orange}\framebox(15,4){}}     

                \put(6.5,2){\color{orange}\framebox(7,3.5){}}  

                \put(91.5,20.5){\color{orange}\framebox(3.5,3.5){}} 
                \put(96,20.5){\color{orange}\framebox(3.5,14){}} 
                \put(5.7,1.7){\color{orange}\framebox(11,14){}}  
                \put(17.5,2){\color{orange}\framebox(3.2,3.5){}} 
            \end{overpic}
            \caption{Unoptimized circuit diagram (partial) compiled from a cyclic permutation matrix.
            Possible annihilations and consolidations are framed.}
            \label{fig:unoptimized-circuit}
        \end{subfigure}

        \vspace{1em}

        \begin{subfigure}{0.8\textwidth}
            \centering
            \begin{overpic}[width=\linewidth,clip,trim=0 2070 0 30]{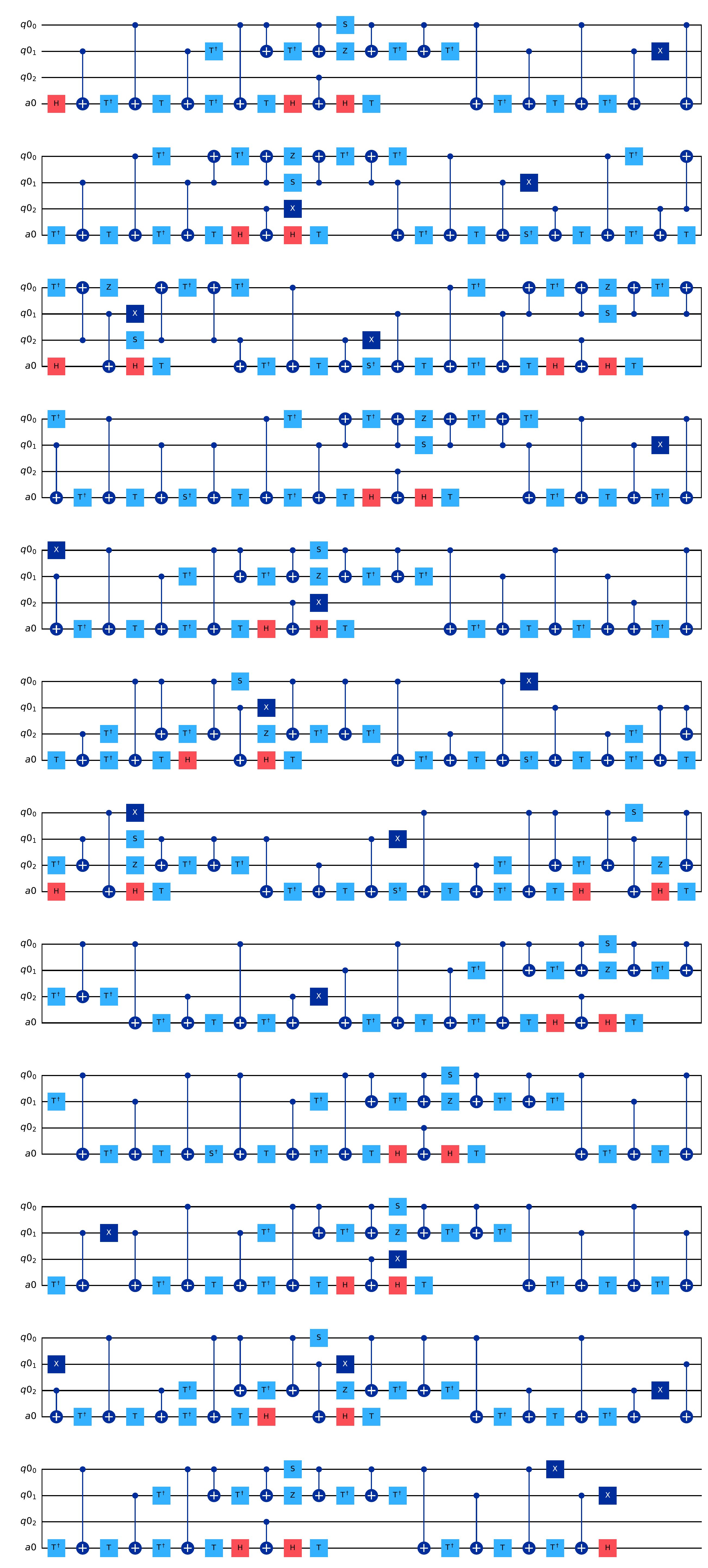}
                \put(47,50.7){\color{orange}\framebox(3.5,3.3){}}  

                \put(47,47){\color{orange}\framebox(3.5,3.3){}}  

                \put(39.7,32){\color{orange}\framebox(3.5,3.5){}}  

                \put(39.7,28){\color{orange}\framebox(3.5,3.5){}}  

                \put(13.7,13.5){\color{orange}\framebox(3.5,3.5){}}  

                \put(17.3,6){\color{orange}\framebox(3.5,3.5){}}  

                \put(73,21){\color{orange}\framebox(3.5,3.5){}} 
            \end{overpic}
            \caption{Same circuit diagram with basic optimization passes.
            Some gates are completely annihilated in the pass.
            The consolidated gates are framed.}
            \label{fig:optimized-circuit}
        \end{subfigure}
        \caption{Side-by-side comparison of compiled circuits before and after optimization.}
        \label{fig:optimized-comparison}
    \end{figure}

    We mentioned the necessity of the recombination steps to cutdown redundant gates and boost efficiency in Section~\ref{sec:high-level-design}.
    To handle this inefficiency, we created the class \texttt{SlidingWindowCombiner} that implements the interface \texttt{Optimizer}
    which is the general interface for optimizers with optimization level and the \texttt{optimize} method.
    The class \texttt{SlidingWindowCombiner} has a list of objects that implement the interface \texttt{WindowOperator} for which
    two concrete classes are implemented: \texttt{AnnihilationOptimizer} and \texttt{ConsolidationOptimizer}
    which detects recombination opportunities among the gates within a window of certain size and performs such optimizations.
    The UML diagram for all the related classes are shown in ~\ref{fig:optimizer-renderer-class-design}.
    After each recombiner takes a pass on the compiled data structure, \texttt{ByteCode},
    the resulting \texttt{ByteCode} is written to a binary file \texttt{.qco}.

    \subsection{Circuit Renderer}\label{subsec:rendering}

    \begin{figure}[tbhp]
        \centering
        \centerline{\includegraphics[width=0.6\textwidth]{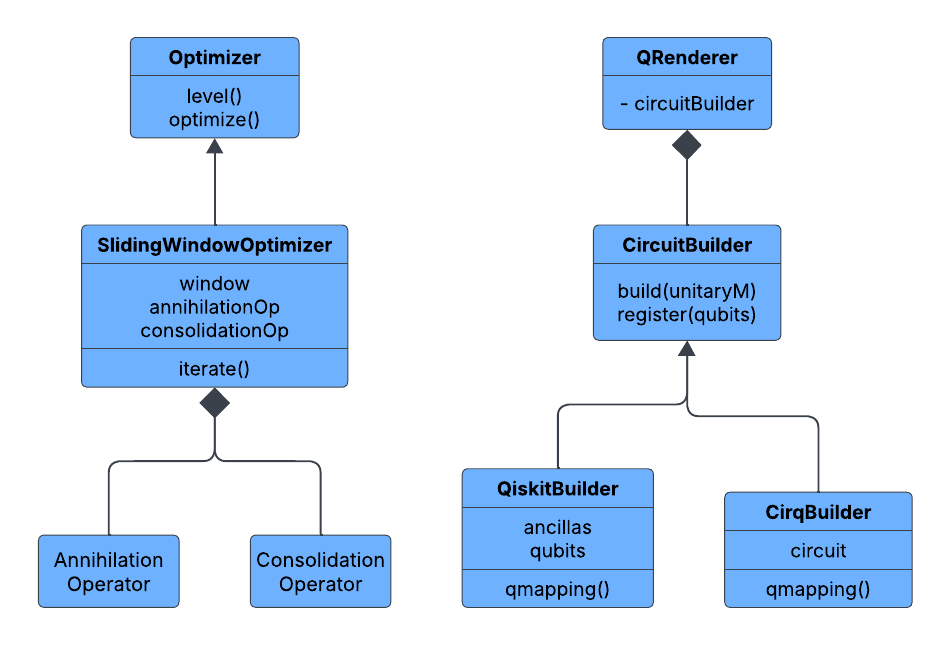}}
        \caption{Renderer and Optimizer class diagrams.
        The relations are \texttt{hasa} and \texttt{isa}.}
        \label{fig:optimizer-renderer-class-design}
    \end{figure}

    As outlined in Section~\ref{sec:high-level-design}, the decomposition process concludes with the creation of the \texttt{.qco} file.
    The rendering stage then uses the \texttt{.qco} file to generate platform-specific quantum circuits.
    The guiding paradigm of the \texttt{Qompiler} is ``Compile once, render multiple times, and run everywhere.''
    We therefore follow the builder pattern to create the class \texttt{QRenderer} and the interface \texttt{CircuitBuilder}.
    The \texttt{QRenderer} class is a concrete class and is responsible for iterating and parsing the IR while delegating
    circuit rendering tasks to \texttt{CircuitBuilder}.
    The \texttt{CircuitBuilder} interface defines the abstract methods needed for circuit building by most known quantum computing platforms.
    \texttt{CircuitBuilder} should be implemented with the platform-specific logic to handle qubits (including ancilla) registration and allocation,
    adding and editing of quantum gates, backend specific circuit optimizations, e.g., based on locality and hardware topology, etc.
    Through polymorphically implementing the \texttt{CircuitBuilder} interface, we realize the cross-platform feature of our system.

    For demonstration purposes, we implemented the \texttt{CircuitBuilder} interface with \texttt{QiskitBuilder} for Qiskit\textregistered\space
    and \texttt{CirqBuilder} for Cirq\texttrademark~\cite{cirq2024,qiskit2023}.
    The transpilation of IR to OpenQASM is realized by first rendering into Qiskit circuit and then export as QASM\@.
    Extending support to a new quantum computing platform is as simple as writing a new implementation of \texttt{CircuitBuilder}.
    The UML diagram for the rendering classes is shown in Figure~\ref{fig:optimizer-renderer-class-design}.

    Hardware-specific optimizations and adaptions may be performed and is recommended at this stage to accommodate for the need of error correction algorithms,
    fine-tune the gate types, swapping qubits to meet connectivity requirements, etc.
    Another important topic at this stage is qubit allocation and mapping based on the connectivity restraints of the physical qubits.
    Proper handling of this is crucial to the performance of the quantum circuit and underpins the feasibility of computation.
    This usually requires highly sophisticated algorithms and heuristics and is beyond the focus of this article.
    We adopt a simplified approach for the proof of concept and leave the study of advanced qubit allocation algorithms for the future.

    \section{Decomposition Processes}\label{sec:decomp}

    \begin{table}[tbhp]
        \centering
        \caption{Main decomposition processes used by \texttt{Qompiler}~\cite{nielsen2010q,dawson2005sk}.}
        \begin{tabularx}{.9\textwidth}{l|X}
            \hline
            \textbf{Decomposition}        & \textbf{Description}                                                                                                                                        \\
            \hline
            \texttt{tl\_decompose}        & Decomposes an arbitrary $n \times n$ unitary matrix into \texttt{2level} unitary matrices.                                                                  \\
            \texttt{gray\_decompose}      & Decomposes a \texttt{2level} unitary matrix into single-target operators in universal gates through gray codes.                                                                \\
            \texttt{ctrl\_decompose}      & Given a controlled gate, breaks its control sequence down to shorter controls, usually with the help of ancilla qubits.                                     \\
            \texttt{euler\_decompose}     & Decomposes a single-qubit gate into Euler rotation operators and optionally a phase shift.                                                                  \\
            \texttt{cliffordt\_decompose} & Given a universal gate, decomposes it into \texttt{Clifford+T} gates.                                                                                       \\
            \texttt{sk\_decompose}        & Based on the Solovay-Kitaev Theorem, approximates a single-qubit gate into \texttt{universal} gates or a subset of it, e.g., the \texttt{Clifford+T} gates. \\
            \hline
        \end{tabularx}
        \label{tab:decomp}
    \end{table}


    \begin{figure}[tbhp]
        \centering
        \includegraphics[width=.8\linewidth]{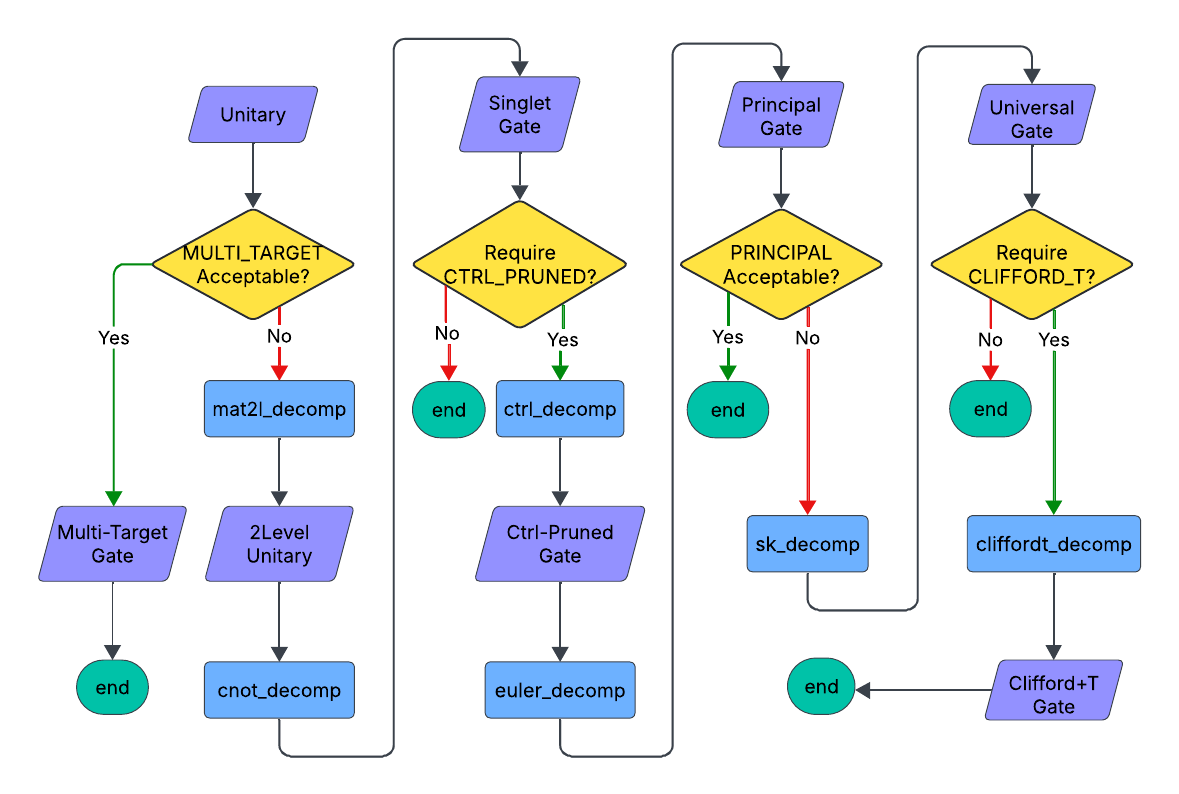}
        \caption{The compilation workflow in details.
        The entry point is \texttt{Unitary}.
        Each parallelogram (purple) represents a stage of the compiled operator.
        Each rhombus (yellow) represents a granularity test.
        Each rectangle (blue) represents a decomposition process.
        }
        \label{fig:workflow}
    \end{figure}

    Starting from an $n$-dimensional unitary matrix, the quantum operators undergoes a number of decomposition processes
    until the resulting quantum gates reach the desired granularity as listed in Table~\ref{tab:terminology}.
    All decomposition processes are orchestrated by the class \texttt{Qompiler} recursively in a depth-first fashion.
    The decomposition processes are listed in Table~\ref{tab:decomp}.
    The granularity of the final quantum gate may be specified in the \texttt{QConfig}.
    The compilation process transfers the dynamics encoded in the input $n \times n$ unitary
    matrix into the quantum circuit, consisting of a network of qubits with quantum gates edges and possibly some classical assets.
    We elaborate the \texttt{Qompiler} decomposition process below.
    Diagram to illustrate the workflow is in Figure~\ref{fig:workflow}.

    If the configuration specifies a granularity of \texttt{MULTI\_TARGET},
    the input unitary matrix does not need any further decomposition.
    The \texttt{Qompiler} will simply assign the necessary \texttt{qspace} and the synthesis process is done.
    Otherwise, \texttt{tl\_decompose} and \texttt{gray\_decompose} will be performed
    to decompose the gates into \texttt{2level} unitary operators, \texttt{CNOT} and \texttt{Singlet} gates, respectively.
    While performing this decomposition, the (virtual) \texttt{qspace} is also allocated which is curated by the class \texttt{QDevice}.
    Next, the \texttt{Qompiler} will determine if the control sequence needs pruning.
    If so, it will perform \texttt{ctrl\_decompose} followed by \texttt{euler\_decompose}
    to break down into \texttt{PRINCIPAL} gates.
    If \texttt{PRINCIPAL} gate types are acceptable, the decomposition process ends.
    Otherwise, the \texttt{sk\_decompose} may be performed to approximate the single-qubit gate
    with a sequence of universal gates.
    (The \texttt{sk\_decompose} will be discussed in Section~\ref{sec:sk-approx} in more details.)
    Then the \texttt{Qompiler} will continue to test if \texttt{Clifford+T} gate is required, and
    if so, the \texttt{cliffordt\_decompose} is performed to conform to the required gate types.

    By the end of decomposition processes, all the gates are guaranteed
    to meet or exceed the configured granularity.
    It is the job of recombiners to fine tune the over-decomposed gates that is explained in Subsection~\ref{subsec:optimization}.

    \section{Solovay--Kitaev Approximation}\label{sec:sk-approx}

    Solovay and Kitaev Theorem (SK) states that any single-qubit operator
    from the \texttt{SU(2)} group can be efficiently approximated using a set of universal gates
    within a chosen error margin $\epsilon$.
    Here efficiency includes both the number of gates of the approximation
    and the runtime to compute it, usually requiring them to be polylogarithmic in terms of $\epsilon$.
    SK is the foundation to quantum computing because the feasibility of efficient expression of
    unitary operators in quantum circuit hangs on this theorem~\cite{kitaev1997quantum,dawson2005sk}.

    The basic idea behind SK approximation is as follows.
    Over the \texttt{SU(2)} operator group, create a point net, e.g., \texttt{SU2Net},
    such that the distance between adjacent points is bound by an error margin $\epsilon_{0}$.
    Each point in \texttt{SU2Net} has a precalculated decomposition,
    which also serves as the zeroth order approximation for its nearest-neighborhood operators.
    Any higher order approximation is obtained by recursively
    calling the function \mintinline{python}{sk_decompose(self, U: NDArray, depth: int)}.
    In each recursion, the group commutator of the residual operator is calculated and used for the next round of approximation
    until it is close enough to the identity operator $I$ or the recursion depth reaches zero.
    The function \texttt{sk\_decompose} is shown in Listing~\ref{lst:sk-decomp}.

    \begin{listing}[h]
        \caption{The SK decomposition function.}
        \label{lst:sk-decomp}
\begin{minted}{python}
def sk_decompose(U: NDArray, n: int) -> Bytecode:
    node, error = su2net.lookup(U)
    if n == 0 or np.isclose(error, 0, atol=rtol, rtol=atol):
        return node
    node = sk_decompose(U, n - 1)
    V, W = gc_decompose(U @ herm(node.data))
    vnode = sk_decompose(V, n - 1)
    wnode = sk_decompose(W, n - 1)
    children = [vnode, wnode, vnode.herm(), wnode.herm(), node]
    data = (children[0].data @ children[1].data @ children[2].data
            @ children[3].data @ children[4].data)
    return Bytecode(data, children=children)
\end{minted}
    \end{listing}

    The caveat of the SK approximation lies in selecting a properly constructed \texttt{SU2Net}.
    For experimentation, we built a few variants of SU2 nets,
    including \texttt{kD-tree}, \texttt{nearest-neighbor} with various metrics.
    Despite our attempts to balance accuracy and runtime, our implementations remains computationally costly.
    Improving this remains a potential direction for future work.

    \section{Results and Discussions}\label{sec:results}
    Initial benchmarks show that the compiler produces correct and consistent quantum circuits across multiple quantum computing platforms.
    These results underscore the compiler’s potential to accelerate quantum software development,
    optimize circuits, and facilitate experimentation across diverse quantum computing platforms.

    To demonstrate \texttt{Qompiler} at work, we compiled a $ 8 \times 8 $ cyclic permutation matrix
    \begin{equation}
        \label{eq:cyclic-permute-mat}
        \begin{bmatrix}
            1 & 0 & 0 & 0 & 0 & 0 & 0 & 0 \\
            0 & 0 & 0 & 0 & 0 & 0 & 0 & 1 \\
            0 & 1 & 0 & 0 & 0 & 0 & 0 & 0 \\
            0 & 0 & 1 & 0 & 0 & 0 & 0 & 0 \\
            0 & 0 & 0 & 1 & 0 & 0 & 0 & 0 \\
            0 & 0 & 0 & 0 & 1 & 0 & 0 & 0 \\
            0 & 0 & 0 & 0 & 0 & 1 & 0 & 0 \\
            0 & 0 & 0 & 0 & 0 & 0 & 1 & 0 \\
        \end{bmatrix}
    \end{equation}
    into universal quantum circuit in Qiskit.
    We have chosen the finest granularity, \texttt{CLIFFORD\_T}, for this demonstration.
    The resulting quantum circuit consists of 279 layers (a Qiskit term referring to groups of related gates),
    and the first dozen are shown in Figure ~\ref{fig:optimized-circuit}.
    We also transpiled our IR can output quantum circuit in OpenQASM format
    and included in Listing~\ref{lst:qasm} for reference.

    \begin{listing}[h]
        \caption{Generated OpenQASM 3.0 code from the cyclic permutation matrix operator.
        The first $40$ lines are shown.}
        \label{lst:qasm}
        \noindent\rule{\linewidth}{0.5pt}
        \vspace{0.5em}
        \begin{minipage}{0.48\linewidth}
\begin{minted}{asm}
OPENQASM 3.0;
include "stdgates.inc";
qubit[3] q0;
qubit[1] a0;
h a0[0];
cx q0[1], a0[0];
tdg a0[0];
cx q0[0], a0[0];
t a0[0];
cx q0[1], a0[0];
tdg a0[0];
cx q0[0], a0[0];
tdg q0[1];
cx q0[0], q0[1];
tdg q0[1];
cx q0[0], q0[1];
s q0[0];
z q0[1];
t a0[0];
h a0[0];
\end{minted}
        \end{minipage}\hfill
        \begin{minipage}{0.48\linewidth}
\begin{minted}{asm}
cx q0[2], a0[0];
h a0[0];
t a0[0];
cx q0[0], q0[1];
tdg q0[1];
cx q0[0], q0[1];
tdg q0[1];
cx q0[0], a0[0];
tdg a0[0];
cx q0[1], a0[0];
t a0[0];
cx q0[0], a0[0];
tdg a0[0];
cx q0[1], a0[0];
x q0[1];
cx q0[0], a0[0];
tdg a0[0];
cx q0[1], a0[0];
t a0[0];
cx q0[0], a0[0];
\end{minted}
        \end{minipage}
        \noindent\rule{\linewidth}{0.5pt}
    \end{listing}

    \begin{figure}[tbhp]
        \centering
        \begin{subfigure}{0.45\textwidth}
            \centering
            \includegraphics[width=\linewidth,clip,trim=0 13100 0 30]{random-unitary}
            \caption{Circuit diagram (partial) in Qiskit\textregistered.}
            \label{fig:qiskit-circuit}
        \end{subfigure}
        \hfill
        \begin{subfigure}{0.45\textwidth}
            \centering
            \includegraphics[width=\textwidth]{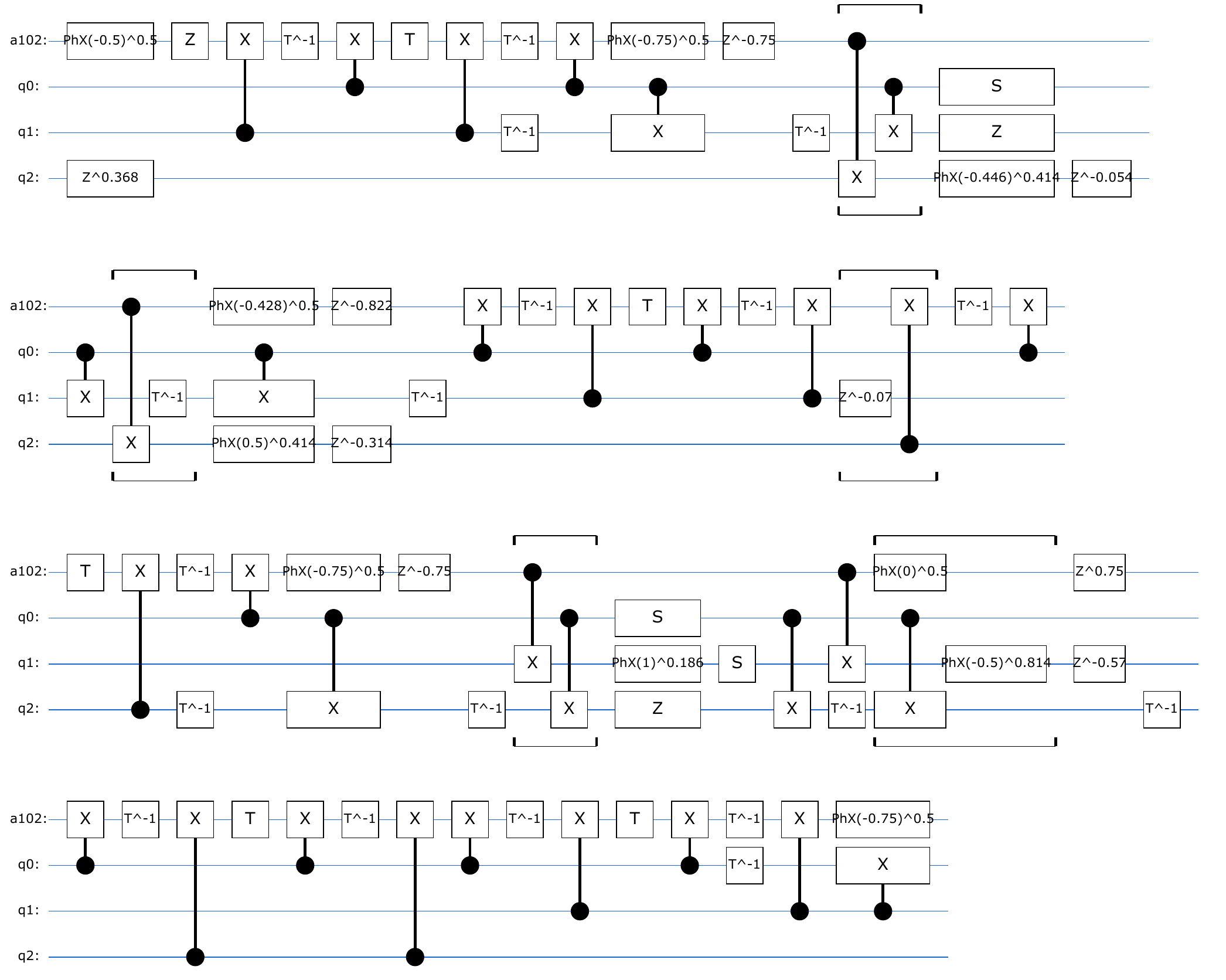}
            \caption{Circuit diagram (partial) in Cirq\texttrademark.}
            \label{fig:cirq-circuit}
        \end{subfigure}
        \caption{Circuit diagrams rendered for different platforms based on the same IR
        that was compiled from a randomly generated unitary matrix and stored in file.
        }
        \label{fig:random-unitary}
    \end{figure}
    As another example, we compiled a randomly generated $8 \times 8$ unitary operator
    \begin{equation}
        \label{eq:random-mat}
        U =
        \begin{bmatrix}
            0.2246+0.3674j & 0.0756+0.3213j & \dots & -0.0534-0.1667j & -0.3321-0.1199j \\
            -0.2123-0.0162j & 0.115+0.4483j & \dots & -0.1772-0.1398j & -0.1267+0.3692j \\
            \vdots & \vdots & \ddots & \vdots & \vdots \\
            0.1554+0.0272j & -0.4272+0.5894j & \dots & 0.1361+0.1934j & 0.3972+0.3205j \\
            -0.3795+0.3577j & -0.0237-0.1054j & \dots & 0.0838-0.4389j & 0.1242-0.0068j
        \end{bmatrix}.
    \end{equation}
    We have chosen a modest granularity, \texttt{PRINCIPAL}, for this demonstration.
    We rendered the resulting quantum circuits in both Qiskit and Cirq, as shown in Figure~\ref{fig:random-unitary}.
    The generated \texttt{QASM} code is similar to Listing~\ref{lst:qasm} which we will not repeat.

    Despite these successes, the \texttt{Qompiler} project has its limitations.
    For example, the project was implemented in Python with numerical libraries, including NumPy and SciPy.
    Therefore, the underlying numerical errors limit the precision of the compiled quantum circuit.
    However, when the Solovay--Kitaev approximation is employed,
    the dominant source of error is the residual approximation error itself.

    As the size of the input increases, practical system limitations arise.
    The program can become computationally intensive for large unitary matrices
    The memory footprint may also become a bottleneck once the number of qubits reaches a certain threshold.

    \section{Conclusion and Future Work}\label{sec:conclusion-and-future-work}
    We presented a configurable, cross-platform quantum compiler, \texttt{Qompiler}, featuring built-in gate traceability.
    Initial benchmarks confirm the correctness and consistency of \texttt{Qompiler}'s
    output across supported platforms, including Qiskit and Cirq.
    It is expected that the pressure on computational resources can be alleviated through distributed computation.
    Future work will be to improve overall efficiency and alleviate the pressure on computational resources through parallelism and distributed system to enable the
    compilation of large input matrices.
    We also plan to add support for additional frameworks, such as AWS Braket and PennyLane,
    with the broader goal of streamlining quantum algorithm development across the heterogeneous quantum
    computing landscape~\cite{braket-zenodo,pennylane}.
    Finally, we would like to implement sophisticated qubit allocation algorithms
    to adapt to given topologies and interaction strategies such as $1$D linear chain,
    $2$D grid, star, all-to-all, etc.
    We welcome collaboration and feedback from the quantum software community.

    \bibliographystyle{unsrt}
    \bibliography{Qompiler}

\begin{thebibliography}{10}

\bibitem{Suzuki1991}
Masuo Suzuki.
\newblock General decomposition theory of ordered exponentials.
\newblock {\em Proceedings of the Japan Academy, Series B}, 69(7):161--166,
  1993.
\newblock Foundational work on Trotter-Suzuki decomposition for exponentials of
  sums of non-commuting operators.

\bibitem{AbramsLloyd1999QPE}
Daniel~S. Abrams and Seth Lloyd.
\newblock Quantum algorithm providing exponential speed increase for finding
  eigenvalues and eigenvectors.
\newblock {\em Phys. Rev. Lett.}, 83:5162--5165, 1999.
\newblock Seminal paper on Quantum Phase Estimation for evaluating eigenvalues
  of unitary operators.

\bibitem{Lloyd1996}
Seth Lloyd.
\newblock Universal quantum simulators.
\newblock {\em Science}, 273(5278):1073--1078, 1996.
\newblock One of the first proposals to implement quantum evolution using
  trotterized sequences.

\bibitem{Chuang1997QPT}
Isaac~L. Chuang and Michael~A. Nielsen.
\newblock Prescription for experimental determination of the dynamics of a
  quantum black box.
\newblock {\em Journal of Modern Optics}, 44(11-12):2455--2467, 1997.

\bibitem{cirq2024}
{Google Quantum AI}.
\newblock Cirq, 2024.
\newblock Online; available at \url{https://quantumai.google/cirq}.

\bibitem{qiskit2023}
H.~Abraham, AduOffei, I.~Y. Akhalwaya, and et~al.
\newblock Qiskit: An open-source framework for quantum computing.
\newblock Zenodo, 2023.

\bibitem{openqasm2}
A.~W. Cross and et~al.
\newblock Open quantum assembly language.
\newblock {\em arXiv preprint arXiv:1707.03429}, 2017.

\bibitem{openqasm3}
A.~W. Cross and et~al.
\newblock Openqasm 3: A broader and deeper quantum assembly language.
\newblock {\em arXiv preprint arXiv:2104.14722}, 2021.

\bibitem{sivarajah2020tket}
Seyon Sivarajah, Silas Dilkes, Alexander Cowtan, Will Simmons, Alec Edgington,
  and Ross Duncan.
\newblock t\ensuremath{|}ket\ensuremath{\rangle}: a retargetable compiler for
  nisq devices.
\newblock {\em Quantum Science and Technology}, 6(1):014003, 2020.

\bibitem{smith2020quilc}
Robert~S. Smith, Eric~C. Peterson, Mark~G. Skilbeck, and Erik~J. Davis.
\newblock An open-source, industrial-strength optimizing compiler for quantum
  programs.
\newblock 2020.

\bibitem{younis2020qfast}
Ed~Younis, Koushik Sen, Katherine Yelick, and Costin Iancu.
\newblock {QFAST}: Quantum synthesis using a hierarchical continuous circuit
  space.
\newblock 2020.

\bibitem{amy2019staq}
Matthew Amy and Vlad Gheorghiu.
\newblock staq -- a full-stack quantum processing toolkit.
\newblock 2019.

\bibitem{Gamma1994DesignPattern}
Erich Gamma, Richard Helm, Ralph Johnson, and John Vlissides.
\newblock {\em Design Patterns: Elements of Reusable Object-Oriented Software}.
\newblock Addison-Wesley Professional Computing Series. Addison-Wesley,
  Reading, MA, 1994.

\bibitem{TracySingh1972}
Derrick~S. Tracy and Rana~P. Singh.
\newblock A new matrix product and its applications in partitioned matrix
  differentiation.
\newblock {\em Statistica Neerlandica}, 26(4):143--157, December 1972.

\bibitem{Hong_2022}
Cheng-Lin Hong et~al.
\newblock Accurate and efficient quantum computations of molecular properties
  using daubechies wavelet molecular orbitals: A benchmark study against
  experimental data.
\newblock {\em PRX Quantum}, 3(2), June 2022.

\bibitem{poulin2014}
David Poulin, M.~B. Hastings, Dave Wecker, Nathan Wiebe, Andrew~C. Doherty, and
  Matthias Troyer.
\newblock The trotter step size required for accurate quantum simulation of
  quantum chemistry.
\newblock 2014.

\bibitem{google_science_2020}
Frank Arute et~al.
\newblock Hartree-fock on a superconducting qubit quantum computer.
\newblock {\em Science}, 369(6507):1084–1089, August 2020.

\bibitem{nielsen2010q}
M.~A. Nielsen and I.~L. Chuang.
\newblock {\em Quantum Computation and Quantum Information}.
\newblock Cambridge University Press, 2010.

\bibitem{dawson2005sk}
C.~M. Dawson and M.~A. Nielsen.
\newblock The solovay-kitaev algorithm.
\newblock {\em arXiv preprint quant-ph/0505030}, 2005.

\bibitem{kitaev1997quantum}
A.~Yu. Kitaev.
\newblock Quantum computations: algorithms and error correction.
\newblock {\em Russian Mathematical Surveys}, 52(6):1191--1249, 1997.

\bibitem{braket-zenodo}
{Amazon Web Services}.
\newblock Amazon braket sdk.
\newblock Zenodo, 2023.

\bibitem{pennylane}
J.~Bergholm and et~al.
\newblock Pennylane: Automatic differentiation of hybrid quantum-classical
  computations.
\newblock {\em arXiv preprint arXiv:1811.04968}, 2018.

\end{thebibliography}

\end{document}